\documentclass[10pt,conference]{IEEEtran}
\usepackage{ifpdf}
\ifpdf
	\usepackage[pdftex]{graphicx}
    \usepackage[update]{epstopdf}
\else
	\usepackage{graphicx}
\fi

\usepackage{amsmath}
\usepackage{amsfonts}
\usepackage{amssymb}
\usepackage{amstext}
\usepackage{latexsym}
\usepackage{color}
\usepackage{cite,setspace}
\usepackage{multirow}
\usepackage{breqn}

\newtheorem{theorem}{\textbf{Theorem}}
\newtheorem{prop}{\textbf{Proposition}}
 
\newtheorem{definition}{\textbf{Definition}}

\begin{document}

\title{Rate Region of the $(4,3,3)$ Exact-Repair \\Regenerating Codes}

\author{\authorblockN{Chao Tian}
\authorblockA{AT\&T Labs-Research, Shannon Laboratory\\
 Florham Park, NJ 07932, USA\\
{\sffamily tian@research.att.com}}}
\maketitle

\textfloatsep=0.21cm
\intextsep=0.21cm
\abovecaptionskip=-0.0cm
\belowcaptionskip=-0.0cm

\abovedisplayskip=0.15cm
\belowdisplayskip=0.15cm

\begin{abstract}
Exact-repair regenerating codes are considered for the case $(n,k,d)=(4,3,3)$, for which a complete characterization of the rate region is provided. This characterization answers in the affirmative the open question whether there exists a non-vanishing gap between the optimal bandwidth-storage tradeoff of the functional-repair regenerating codes ({\em i.e.,} the cut-set bound) and that of the exact-repair regenerating codes. The converse proof relies on the existence of symmetric optimal solutions. For the achievability, only one non-trivial corner point of the rate region needs to be addressed, for which an explicit binary code construction is given. 
\end{abstract}

\section{Introduction}
Dimakis {\em et al.} \cite{Dimakis:10} proposed the framework of regenerating codes to address the tradeoff between the storage and repair bandwidth in erasure-coded distributed storage systems. In this framework, the overall system consists of $n$ storage nodes situated in different network locations, each with $\alpha$ units of data, and the content is coded in such a way that by accessing any $k$ of these $n$ storage nodes,  the full data content of $B$ units can be completely recovered. When a node fails, a new node may access any $d$ remaining nodes for $\beta$ units of data each, in order to regenerate a new data node. 

The main result in \cite{Dimakis:10} is for the so-called functional-repair case, where the regenerating process only needs to guarantee that the regenerated node can serve the same purpose as the lost node, {\em i.e.}, data reconstruction using any $k$ nodes, and being able to help regenerate new data nodes to replace subsequently failed nodes. It was shown that this problem can be cleverly converted to a network multicast problem, and the celebrated result on network coding \cite{Yeung:00} can be applied directly to provide a complete characterization of the optimal bandwidth-storage tradeoff. Furthermore, linear network codes \cite{Yeung:03} are sufficient to achieve this optimal performance.

The decoding and repair rules for functional-repair regenerating codes may evolve as nodes are repaired, which increases the overhead of the system. Moreover, functional-repair does not guarantee systematic format storage, which is an important requirement in practice. For this reason, exact-repair regenerating codes have received considerable attention recently \cite{RashmiShah:11,RashmiShah:12:1,RashmiShah:12:2,Cadambe:11}, where the regenerated data need to be exactly the same as that stored in the failed node. 

The optimal bandwith-storage tradeoff for the functional-repair case can clearly serve as an outer bound for the exact-repair case. There also exist code constructions for the two extreme cases, {\em i.e.}, the minimum storage regenerating (MSR) point \cite{RashmiShah:11,RashmiShah:12:2,Cadambe:11}, or the minimum bandwidth regenerating (MBR) point \cite{RashmiShah:11,RashmiShah:12:1}, and the aforementioned outer bound is in fact achievable at these two extreme points. Also relevant is the fact that symbol extensions are necessary for linear codes to achieve the MSR point for some parameter range \cite{RashmiShah:12:2}, however  the MSR point can indeed be asymptotically (in $B$) achieved by linear codes for all the parameter range \cite{Cadambe:11}. It was also shown in \cite{RashmiShah:12:1} that other than the two extreme points and a segment close to the MSR point, the majority the functional repair outer bound is in fact not strictly achievable by exact-repair regenerating codes. 

The non-achievability result reported in \cite{RashmiShah:12:1} was proved by contradiction, {\em i.e.},  a contradiction will occur if one supposes that an exact-repair code operates {\em strictly} on the optimal functional-repair tradeoff curve. However, it is not clear whether this contradiction is caused by the functional-repair outer bound being only asymptotically achievable, or caused by the existence of a non-vanishing gap between the optimal tradeoff of exact-repair codes and the functional-repair outer bound. In fact, the necessity of symbol extension proved in \cite{RashmiShah:12:2} and the asymptotically optimal construction given in \cite{Cadambe:11} may be interpreted as suggesting that the former is true. 

In this work, we focus on the simplest case of exact-repair regenerating codes, {\em i.e.}, when $(n,k,d)=(4,3,3)$, for which the rate region has not been completely characterized previously. A complete characterization of the rate region is provided for this case, which shows that indeed there exists a non-vanishing gap between the optimal tradeoff of the exact-repair codes and that of the functional-repair codes. As in many open information theoretical problems, the difficulty lies in finding good outer bounds, particularly in this problem with a large number of regenerating and reconstruction requirements. We rely on a computer-aided proof (CAP) approach and take advantage of the symmetry in the problem to reduce the computation complexity. This approach builds upon Yeung\rq{}s linear programming (LP) framework \cite{Yeung:book}, but instead of only machine-proving whether an information theoretic bound is true or not as in \cite{Yeung:book}, we further use a secondary optimization procedure to find an \textit{explicit information theoretic proof}, which, after some amount of machine-to-human translation, is presented here. As of our knowledge, this is the first time that the LP framework is meaningfully applied to a non-trivial information theoretic problem, which leads to a complete solution. 
Although this CAP approach may be of independent interest by itself, due to space constraint we focus in this paper on establishing the rate region, and only briefly discuss the CAP approach and leave the details to another work. 

The rest of the paper is organized as follows. In Section \ref{sec:definition}, we provide a formal definition of the problem and review briefly the functional-repair outer bound.  The main result of this paper is given in Section \ref{sec:main}. The code construction for the achievability part is given in Section \ref{sec:achievability}, and the converse is proved in Section \ref{sec:converse}. Section \ref{sec:conclusion} concludes the paper.

\section{Problem Definition}
\label{sec:definition}


\subsection{Exact-Repair Regenerating Codes}
A $(4,3,3)$ exact-repair regenerating code is formally defined as follows, where the notation $I_n$ is used to denote the set $\{1,2,\ldots,n\}$, and $|A|$ is used for the cardinality of a set $A$.
\begin{definition}
An $(N,K_d,K)$ exact-repair regenerating code for the $(4,3,3)$ case consists of $4$ encoding function $f^E_i(\cdot)$,  $4$ decoding functions $f^D_{A}(\cdot)$, $12$ repair encoding functions $F^{E}_{i,j}(\cdot)$,  and $4$ repair decoding functions $F^{D}_{i}(\cdot)$, where
\begin{align*}
f^E_i:I_N\rightarrow I_{K_d},\quad i\in I_4,
\end{align*}
each of which maps the message $m\in I_N$ to one piece of coded information, 
\begin{align*}
f^D_{A}:I_{K_d}\times I_{K_d}\times I_{K_d}\rightarrow I_N,\quad A\subset {I}_4\quad \mbox{and}\quad |A|=3,
\end{align*}
each of which maps $3$ pieces of coded information stored on a set $A$ of nodes to the original message, 
\begin{align*}
F^{E}_{i,j}:I_{K_d}\rightarrow I_{K},\quad j\in I_4,\quad\mbox{and}\quad i\in I_4\setminus \{j\},
\end{align*}
each of which maps a piece of coded information at node $i$ to an index that will be made available to reconstruct the data at node $j$, and
\begin{align*}
F^{D}_{j}:{I}_{K}\times{I}_{K}\times{I}_{K} \rightarrow {I}_{K_d},\quad \quad j\in{I}_4,
\end{align*}
each of which maps 3 such indices from the helper nodes to reconstruct the information stored at the failed node.  The functions must satisfy the data reconstruction conditions
\begin{align*}
f_{{A}}^D\left(\prod_{i\in{A}}f^E_i(m)\right)=m,\quad m\in{I}_N,\,\,{A}\subset {I}_4\,\, \mbox{and}\,\, |{A}|=3,
\end{align*}
and the repair conditions
\begin{align*}
F^D_{j}\left(\prod_{i\in{I_n\setminus \{j\}}}F^E_{i,j}\left(f^E_i(m)\right)\right)=f^E_j(m),\,\, m\in{I}_N,\quad j\in{I}_4.
\end{align*}
\end{definition}

\begin{definition}
A normalized bandwidth-storage pair $(\bar{\alpha},\bar{\beta})$ is said to be $(4,3,3)$ exact-repair achievable if for any $\epsilon>0$ there exists an $(N,K_d,K)$ exact-repair regenerating code such that
\begin{align}
\bar{\alpha}+\epsilon\geq \frac{\log K_d}{\log N},\quad
\bar{\beta}+\epsilon\geq \frac{\log K}{\log N}.
\end{align}
The collection of all the achievable $(\bar{\alpha},\bar{\beta})$ pairs is the achievable region $\mathcal{R}$ of the $(4,3,3)$ exact-repair regenerating codes.
\end{definition}

The quantity $\epsilon$ in the definition is introduced to include the case when the storage-bandwidth tradeoff may be approached asymptotically, {\em e.g.}, the case considered in \cite{Cadambe:11}.

\subsection{Some Further Notation}
In order to derive the outer bound, it is convenient to write the reconstruction and regenerating conditions in the form of entropy constraints. For this purpose, some further notation is introduced here, which is largely borrowed from \cite{RashmiShah:12:1}.
Let us denote the message random variable as $M$, which is uniformly distributed in the set $I_N$. Define
\begin{align}
W_i=f^E_i(M),\quad S_{i,j}=F^E_{i,j}\left(f^E_i(M)\right).
\end{align}
Thus we have the following random variables in the set $\mathcal{W}\cup\mathcal{S}$
\begin{align}
\mathcal{W}=&\{W_1,W_2,W_3,W_4\},\\
\mathcal{S}=&\{S_{1,2},S_{1,3},S_{1,4},S_{2,1},S_{2,3},S_{2,4},\nonumber\\
&\quad S_{3,1},S_{3,2},S_{3,4},S_{4,1},S_{4,2},S_{4,3}\}.
\end{align}
The reconstruction requirement thus implies that
\begin{align}
H(\mathcal{W}\cup\mathcal{S}|\mathcal{A})=0,\quad \mbox{any } \mathcal{A}\subseteq \mathcal{W}: |\mathcal{A}|=3.
\label{eqn:reconstruction}
\end{align}
The regenerating requirement implies that
\begin{align}
H(S_{i,j}|W_i)=0,\quad j\in I_4,\quad i\in I_4\setminus \{j\},
\label{eqn:regeneratingencoding}
\end{align}
and 
\begin{align}
H(W_j|\{S_{i,j}\in \mathcal{S}: i\in I_n\setminus\{j\}\})=0,\quad \mbox{any } j\in I_4. 
\label{eqn:regenerating}
\end{align}
Because the message $M$ has a uniform distribution, we also have that
\begin{align}
H(\mathcal{W}\cup\mathcal{S})=H(M)=\log N\triangleq B,
\label{eqn:totalinfo}
\end{align}
which is strictly larger than zero. Note that together with (\ref{eqn:reconstruction}), this implies that
\begin{align}
H(\mathcal{A})=B,\quad \mbox{any } \mathcal{A} \mbox{ such that } |\mathcal{A}\cap\mathcal{W}|\geq 3.
\label{eqn:totalinfo2}
\end{align}
The symmetric storage requirement can be written as
\begin{align}
H(W_i)\leq \log K_d \triangleq \alpha,\quad W_i\in \mathcal{W},
\label{eqn:alpha}
\end{align}
and the regenerating bandwidth constraint can be written as
\begin{align}
H(S_{i,j})\leq \log K \triangleq \beta,\quad S_{i,j}\in \mathcal{S}.
\label{eqn:beta}
\end{align}
The above constraints (\ref{eqn:reconstruction})-(\ref{eqn:beta}) are the constraints that need to be satisfied by any exact-repair regenerating codes. 

\begin{figure}[tcb]
  \centering
  \includegraphics[width=.8\linewidth]{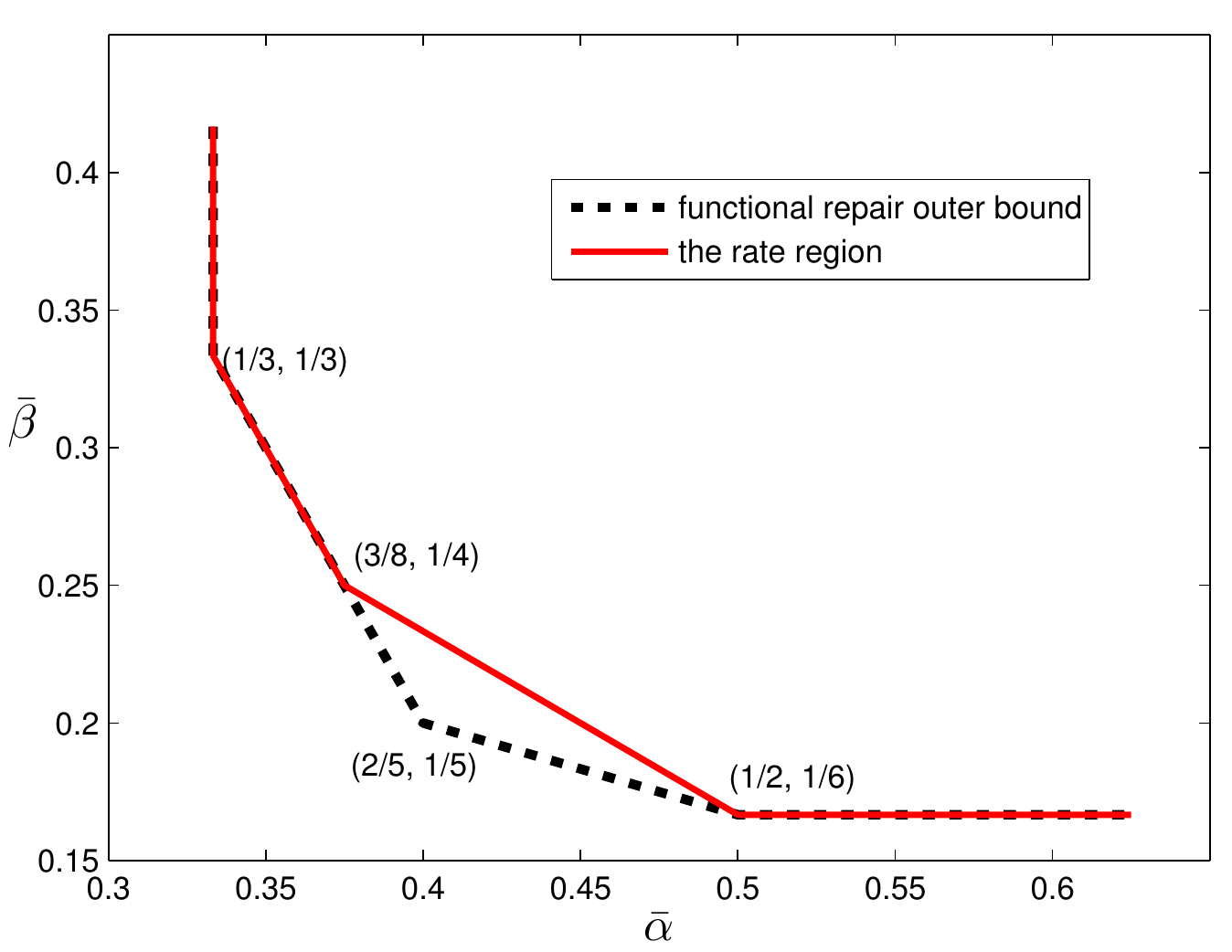}
  \caption{The functional-repair outer bound and the rate-region $\mathcal{R}$.\label{fig:bound433}}
\end{figure}

\subsection{Review of Functional-Repair Outer Bound}
\label{sec:review}
The optimal tradeoff for functional-repair regenerating codes was given by Dimakis {\em et al.} \cite{Dimakis:10}, which provides an outer bound for the exact-repair case. The bound has the following form in our notation for the $(4,3,3)$ case (see Fig. \ref{fig:bound433})
\begin{align}
\sum_{i=0}^2\min(\bar{\alpha},(3-i)\bar{\beta})\geq 1.
\end{align}
It is not difficult to show that it can be rewritten as the following four simultaneous linear bounds
\begin{align}
3\bar{\alpha}\geq 1,\quad 2\bar{\alpha}+\bar{\beta}\geq 1,\quad
\bar{\alpha}+3\bar{\beta}\geq 1,\quad 6\bar{\beta}\geq 1.\label{eqn:frouterbound}
\end{align}
The MSR point for this case is $(\bar{\alpha},\bar\beta)=(\frac{1}{3},\frac{1}{3})$, and the MBR point is  $(\bar{\alpha},\bar\beta)=(\frac{1}{2},\frac{1}{6})$.

\section{Main Result}
\label{sec:main}

The main result in this paper is the following theorem.
\begin{theorem}
The rate region $\mathcal{R}$ of the $(n,k,d)=(4,3,3)$ exact-repair regenerating codes is given by
\begin{align*}
3\bar{\alpha}\geq 1,\quad 2\bar{\alpha}+\bar{\beta}\geq 1,\quad 4\bar{\alpha}+6\bar{\beta}\geq 3,\quad 6\bar{\beta}\geq 1.
\end{align*}
\end{theorem}

This rate region is also depicted in Fig. \ref{fig:bound433}.

\section{Achievability Proof}
\label{sec:achievability}

The rate region $\mathcal{R}$ has three corner points, and thus we only need to show that these three points are all achievable. The MSR point $(\bar{\alpha},\bar\beta)=(\frac{1}{3},\frac{1}{3})$ is simply achieved by any $(4,3)$  MDS code, such as the binary systematic code with a single parity check bit. The MBR point $(\bar{\alpha},\bar\beta)=(\frac{1}{2},\frac{1}{6})$ is also easily obtained by using the repair-by-transfer code construction in \cite{RashmiShah:12:1}, which in this case reduces to a simple replication coding. It thus only remains to show that the point  $(\bar{\alpha},\bar\beta)=(\frac{3}{8},\frac{1}{4})$ is also achievable. 

Next we shall give a construction for a binary $(4,3,3)$ code with $\alpha=3$, $\beta=2$ and $B=8$, which indeed achieves this operating point. The code is illustrated in Table \ref{table:code}, where (and in the remainder of this section) the addition $+$ is in the binary field. Here $x_i,y_i,z_i,t_i$ are the systematic bits, $i=1,2$, and the remaining bits are the parity bits.

First note that the construction is circularly symmetric, and thus without loss of generality, we only need to consider the case when node 1 fails. If it can be shown that when node 2, 3, 4 each contribute two bits, node 1 can be reconstructed, which also implies that the complete data can be recovered using only node 2, 3 and 4, then the proof is complete. This can indeed be done using the combination shown in Table \ref{table:repair}.

Upon receiving these six bits  in Table \ref{table:repair}, the new node can form the following combinations
\begin{align*}
x_1+x_2&\quad+\quad y_1+y_2+z_1+\mathbf{z_2}+t_2\\
x_2&\quad+\quad y_1+y_2+z_1+z_2+t_1+\mathbf{t_1+t_2}\\
x_1&\quad+\quad \mathbf{y_1}+y_2+z_1+z_2+t_2,
\end{align*}
where the first combination is formed by using the second bit from node 2 and the first bit from node 3 (shown in bold), and the other combinations can be formed similarly. 
In the binary field, this is equivalent to having
\begin{align}
x_1+x_2&\quad+\quad y_1+y_2+z_1+z_2+t_2\label{eqn:first}\\
x_2&\quad+\quad y_1+y_2+z_1+z_2+t_2\label{eqn:second}\\
x_1&\quad+\quad y_1+y_2+z_1+z_2+t_2,\label{eqn:third}
\end{align} 
and it is seen that $x_1$ can be recovered by simply taking the difference between (\ref{eqn:first}) and (\ref{eqn:second}), and similarly $x_2$ can be recovered by taking the difference between (\ref{eqn:first}) and (\ref{eqn:third}). Note further that the third bit stored in node 1 is simply the summation of the first bits contributed from node 2, 3, and 4 in Table \ref{table:repair}. The proof is thus complete. \QED

\begin{table}
\centering
\caption{A $(4,3,3)$ code for  $(\bar{\alpha},\bar\beta)=(\frac{3}{8},\frac{1}{4})$.}
\label{table:code}
\begin{tabular}{|c|c|c|c|}
\hline
&first bit&second bit&third bit\\\hline
node 1&$x_1$&$x_2$&$y_1+z_2+t_1+t_2$\\\hline
node 2&$y_1$&$y_2$&$z_1+t_2+x_1+x_2$\\\hline
node 3&$z_1$&$z_2$&$t_1+x_2+y_1+y_2$\\\hline
node 4&$t_1$&$t_2$&$x_1+y_2+z_1+z_2$\\\hline
\end{tabular}
\end{table}

\begin{table}
\centering
\caption{Repair contributions when node 1 fails.}
\label{table:repair}
\begin{tabular}{|c|c|c|}
\hline
&first bit&second bit\\\hline
node 2&$y_1$&$z_1+t_2+x_1+x_2+y_1+y_2$\\\hline
node 3&$z_2$&$t_1+x_2+y_1+y_2+z_1+z_2$\\\hline
node 4&$t_1+t_2$&$x_1+y_2+z_1+z_2+t_2$\\\hline
\end{tabular}
\end{table}

\section{Converse Proof}
\label{sec:converse}

It is clear that we only need to prove the following bound
\begin{align}
4\bar{\alpha}+6\bar{\beta}\geq 3,
\end{align}
because the other bounds in the main theorem can be obtained from the outer bound (\ref{eqn:frouterbound}). We first give an instrumental result regarding the symmetry of the optimal solution.

\subsection{A Symmetry Reduction}

\begin{definition}
A permutation $\pi$ on the set $I_4$ is a one-to-one mapping $\pi:{I}_4\rightarrow {I}_4$. The collections of all permutations is denoted as $\Pi$.
\end{definition}

Any given permutation $\pi$ correspondingly maps a random variable $W_i$ to $W_{\pi(i)}$.  Any subset of $\mathcal{W}$, e.g., $\mathcal{A}\subseteq \mathcal{W}$, is thus mapped to another set of random variables, denoted as $\pi(\mathcal{A})$. For example, the permutation $\pi(1)=2$, $\pi(2)=3$, $\pi(3)=1$ and $\pi(4)=4$ will map the set of random variables $\mathcal{A}=\{W_1,W_4\}$ to $\pi(\mathcal{A})=\{W_2,W_4\}$. Similarly a random variable $S_{i,j}$ will be mapped to $S_{\pi(i),\pi(j)}$, and for any subset of $\mathcal{S}$, we use a similar notation as for the case of $\mathcal{W}$.

\begin{definition}
An $(N,K_d,K)$ exact-repair regenerating code is said to induce a symmetric entropic vector if for any sets $\mathcal{A}\subseteq \mathcal{S}$ and $\mathcal{B}\subseteq\mathcal{W}$ and any permutation $\pi\in\Pi$, 
\begin{align}
H(\mathcal{A},\mathcal{B})=H(\pi(\mathcal{A}),\pi(\mathcal{B})).
\end{align}
\end{definition}

\begin{definition}
A normalized bandwidth-storage pair $(\bar{\alpha},\bar{\beta})$ is said to be entropy-symmetrically $(4,3,3)$ exact-repair achievable if for any $\epsilon>0$ there exists an $(N,K_d,K)$ exact-repair regenerating code which induces a symmetric entropic vector such that
\begin{align}
\bar{\alpha}+\epsilon\geq \frac{\log K_d}{\log N},\quad
\bar{\beta}+\epsilon\geq \frac{\log K}{\log N}.
\end{align}
The collection of all such $(\bar{\alpha},\bar{\beta})$ pairs is the symmetrically achievable region $\mathcal{R}^*$ of the $(4,3,3)$ exact-repair regenerating codes.
\end{definition}

With the above definition, it is not difficult to see the following proposition is true.
\begin{prop}
\label{prop:symmetry}
For $(n,k,d)=(4,3,3)$ exact-repair regenerating codes $\mathcal{R}=\mathcal{R}^*$.
\end{prop}

Clearly the inclusion $\mathcal{R}^*\subseteq\mathcal{R}$ is true.  For the other direction, we can simply invoke a time-sharing (or more precisely here, space-sharing) argument among all possible permutations; the proof details are thus omitted. 

\subsection{Converse Proof of Theorem 1}

Because of the equivalence in Proposition \ref{prop:symmetry}, without loss of generality we can limit ourselves to only codes that induce symmetrical entropy vectors. We first write
\begin{align}
&8\alpha + 12\beta \nonumber\\
\geq& 4H(S_{3,1}S_{2,1}W_4)+4H(S_{3,2}W_4)\nonumber\\
\geq& 4H(S_{3,1}S_{2,1}W_4)+4H(S_{3,2}W_4)\nonumber\\
&-2I(S_{2,1};W_3|W_4)-2I(W_3;W_4|S_{3,2}S_{3,4}S_{2,4})\nonumber\\
=&4H(S_{3,1}S_{2,1}W_4)+2H(W_4)+2H(W_3W_4S_{2,1})\nonumber\\
&-2H(W_3W_4)+2H(S_{3,2}S_{3,4}S_{2,4})\nonumber\\
&+2H(W_3W_4S_{2,4}S_{3,2}S_{3,4})-2H(W_4S_{3,2}S_{3,4}S_{2,4})\nonumber\\
=&4H(S_{3,1}S_{2,1}W_4)+2H(W_4)+2B\nonumber\\
&-2H(W_3W_4)+2H(S_{3,2}S_{3,4}S_{2,4})\nonumber\\
&+2H(W_3W_4S_{2,4})-2H(W_4S_{3,2}S_{3,4}S_{2,4})
\end{align}
where the first inequality is by (\ref{eqn:alpha}) and (\ref{eqn:beta}), the first equality is by expanding the mutual information terms and using 
\begin{align}
H(S_{3,2}S_{3,4}S_{2,4}W_3)=H(S_{2,4}W_3)
\end{align}
implied by (\ref{eqn:regeneratingencoding}) and then the symmetry
\begin{align}
H(S_{3,2}W_4)=H(S_{2,4}W_3)=H(S_{2,1}W_4),
\end{align}
and the second equality is by (\ref{eqn:regeneratingencoding}), (\ref{eqn:regenerating}) and (\ref{eqn:totalinfo2}) on the third term
\begin{align}
H(W_3W_4S_{2,1})&\stackrel{(\ref{eqn:regeneratingencoding})}{=}H(W_3W_4S_{2,1}S_{3,1}S_{4,1})\nonumber\\
&\stackrel{(\ref{eqn:regenerating})}{=}H(W_1W_3W_4S_{2,1}S_{3,1}S_{4,1})\stackrel{(\ref{eqn:totalinfo2})}{=}B,
\end{align}
and (\ref{eqn:regeneratingencoding}) on the sixth term.
For notational simplicity,  we shall write from here on (s), (\ref{eqn:reconstruction}), (\ref{eqn:regeneratingencoding}), (\ref{eqn:regenerating}) and (\ref{eqn:totalinfo2}) on top of the equalities in the derivation to signal the reasons for the equalities, {\em i.e.,} by the symmetry, or by equations (\ref{eqn:reconstruction}), (\ref{eqn:regeneratingencoding}), (\ref{eqn:regenerating}) and (\ref{eqn:totalinfo2}), respectively. 

Note that 
\begin{align}
&H(S_{3,1}S_{2,1}W_4)+H(S_{3,2}S_{3,4}S_{2,4})\nonumber\\
\stackrel{(s)}{=}&H(S_{3,4}S_{2,4}W_1)+H(S_{3,2}S_{3,4}S_{2,4})\nonumber\\
=&H(W_1|S_{3,4}S_{2,4})+H(S_{3,2}|S_{3,4}S_{2,4})+2H(S_{3,4}S_{2,4})\nonumber\\
\geq& H(W_1S_{3,2}|S_{3,4}S_{2,4})+2H(S_{3,4}S_{2,4})\nonumber\\
=&H(W_1S_{3,2}S_{3,4}S_{2,4})+H(S_{3,4}S_{2,4})\nonumber\\
\stackrel{(\ref{eqn:regeneratingencoding},\ref{eqn:regenerating})}{=}&H(W_1W_2W_4S_{3,2}S_{3,4})+H(S_{3,4}S_{2,4})\nonumber\\
\stackrel{ (\ref{eqn:totalinfo2})}=&B+H(S_{3,4}S_{2,4}),
\end{align}
where, to be more precise, the last but one equality is because
\begin{align}
&H(W_1S_{3,2}S_{3,4}S_{2,4})
\stackrel{(\ref{eqn:regeneratingencoding})}{=}H(W_1S_{1,4}S_{3,2}S_{3,4}S_{2,4})\nonumber\\
&\stackrel{(\ref{eqn:regenerating})}{=}H(W_1W_4S_{1,4}S_{3,2}S_{3,4}S_{2,4})\nonumber\\
&\stackrel{(\ref{eqn:regeneratingencoding})}{=}H(W_1W_4S_{1,2}S_{4,2}S_{1,4}S_{3,2}S_{3,4}S_{2,4})\nonumber\\
&\stackrel{(\ref{eqn:regenerating})}{=}H(W_1W_2W_4S_{1,2}S_{4,2}S_{1,4}S_{3,2}S_{3,4}S_{2,4})\nonumber\\
&=H(W_1W_2W_4S_{3,2}S_{3,4}).
\end{align}
It follows that
\begin{align}
&8\alpha + 12\beta \nonumber\\
\geq& 2H(S_{3,1}S_{2,1}W_4)+2H(W_4)+4B\nonumber\\
&-2H(W_3W_4)+2H(W_3W_4S_{2,4})\nonumber\\
&-2H(W_4S_{3,2}S_{3,4}S_{2,4})+2H(S_{3,4}S_{2,4})\nonumber\\
\stackrel{(s)}{=}&2H(S_{3,1}S_{2,1}W_4)+2H(W_4)+4B\nonumber\\
&-2H(W_3W_4)+2H(W_3W_4S_{2,4})\nonumber\\
&-2H(W_4S_{3,2}S_{3,4}S_{2,4})+2H(S_{3,1}S_{2,1})\nonumber\\
{\geq}& 2H(S_{3,1}S_{2,1}W_4)+2H(S_{3,1}S_{2,1}W_4)+4B\nonumber\\
&-2H(W_3W_4)+2H(W_3W_4S_{2,4})\nonumber\\
&-2H(W_4S_{3,2}S_{3,4}S_{2,4})\nonumber\\
=&4H(S_{3,1}S_{2,1}W_4)+4B-2H(W_3W_4)\nonumber\\
&+2H(W_3W_4S_{2,4})-2H(W_4S_{3,2}S_{3,4}S_{2,4}).
\end{align}

Notice that
\begin{align}
&H(W_3W_4S_{2,4})-H(W_3W_4)\nonumber\\
&=H(S_{2,4}|W_3W_4)\nonumber\\
&\geq H(S_{2,4}|S_{1,3}W_3W_4)\nonumber\\
&=H(S_{2,4}S_{1,3}W_3W_4)-H(S_{1,3}W_3W_4),
\end{align}
then we can further write
\begin{align}
&8\alpha + 12\beta\nonumber\\
\geq& 4H(S_{3,1}S_{2,1}W_4)+4B-2H(S_{1,3}W_3W_4)\nonumber\\
&+2H(S_{2,4}S_{1,3}W_3W_4)-2H(W_4S_{3,2}S_{3,4}S_{2,4})\nonumber\\
\stackrel{(\ref{eqn:regeneratingencoding},\ref{eqn:regenerating})}{=}&2H(S_{3,1}S_{2,1}W_4)+2H(S_{3,1}S_{2,1}W_4W_1)\nonumber\\
&+4B-2H(S_{1,3}W_3W_4)\nonumber\\
&+2H(S_{2,4}S_{1,3}W_3W_4)-2H(W_4S_{3,2}S_{3,4}S_{2,4})\nonumber\\
\stackrel{(s)}{=}&2H(S_{3,1}S_{2,1}W_4)+2H(S_{2,3}S_{1,3}W_3W_4)\nonumber\\
&+4B-2H(S_{1,3}W_3W_4)\nonumber\\
&+2H(S_{2,4}S_{1,3}W_3W_4)-2H(W_4S_{3,2}S_{3,4}S_{2,4}).
\end{align}
However
\begin{align}
&H(S_{2,3}S_{1,3}W_3W_4)+H(S_{2,4}S_{1,3}W_3W_4)\nonumber\\
&-H(S_{1,3}W_3W_4)\nonumber\\
=&H(S_{2,3}|S_{1,3}W_3W_4)+H(S_{2,4}|S_{1,3}W_3W_4)\nonumber\\
&+H(S_{1,3}W_3W_4)\nonumber\\
\geq& H(S_{2,3}S_{2,4}|S_{1,3}W_3W_4)+H(S_{1,3}W_3W_4)\nonumber\\
=&H(S_{2,3}S_{2,4}S_{1,3}W_3W_4),
\end{align}
which leads to 
\begin{align*}
&8\alpha + 12\beta
\geq2H(S_{3,1}S_{2,1}W_4)+4B\nonumber\\
&\qquad+2H(S_{2,3}S_{2,4}S_{1,3}W_3W_4)-2H(W_4S_{3,2}S_{3,4}S_{2,4})\nonumber\\
&\qquad\stackrel{(\ref{eqn:regenerating})}{\geq} 2H(S_{3,1}S_{2,1}W_4)+4B\nonumber\\
&\qquad+2H(S_{2,3}S_{2,4}S_{1,3}W_3W_4)-2H(S_{3,2}S_{3,4}S_{2,4}S_{1,4}),
\end{align*}
where in the above step labeled (\ref{eqn:regenerating}) we have first used the fact that $H(W_4S_{3,2}S_{3,4}S_{2,4})\leq H(W_4S_{3,2}S_{3,4}S_{2,4}S_{1,4})$. 
Because
\begin{align}
&2H(S_{2,3}S_{2,4}S_{1,3}W_3W_4)\nonumber\\
\stackrel{(s)}{=}&H(S_{2,4}S_{2,3}S_{1,4}W_3W_4)+H(S_{3,4}S_{3,2}S_{1,4}W_2W_4)\nonumber\\
\stackrel{(\ref{eqn:regeneratingencoding})}{=}&H(W_3W_4S_{1,4}S_{2,4}S_{3,4}S_{2,3}S_{3,2})\nonumber\\
&+H(W_2W_4S_{1,4}S_{2,4}S_{3,4}S_{2,3}S_{3,2})\nonumber\\
=&H(W_3W_4|S_{1,4}S_{2,4}S_{3,4}S_{2,3}S_{3,2})\nonumber\\
&+H(W_2W_4|S_{1,4}S_{2,4}S_{3,4}S_{2,3}S_{3,2})\nonumber\\
&+2H(S_{1,4}S_{2,4}S_{3,4}S_{2,3}S_{3,2})\nonumber\\
\geq& H(W_2W_3W_4S_{1,4}S_{2,4}S_{3,4}S_{2,3}S_{3,2})\nonumber\\
&+H(S_{1,4}S_{2,4}S_{3,4}S_{2,3}S_{3,2})\nonumber\\
\stackrel{(\ref{eqn:totalinfo2})}=&B+H(S_{1,4}S_{2,4}S_{3,4}S_{2,3}S_{3,2}),
\end{align}
we can now write
\begin{align}
&8\alpha + 12\beta\nonumber\\
\geq& 2H(S_{3,1}S_{2,1}W_4)+5B-2H(S_{3,2}S_{3,4}S_{2,4}S_{1,4})\nonumber\\
&+H(S_{1,4}S_{2,4}S_{3,4}S_{2,3}S_{3,2})\nonumber\\
\stackrel{(s)}{=}&H(S_{3,1}S_{2,1}W_4)+H(S_{1,4}S_{2,4}W_3)+5B\nonumber\\
&-2H(S_{3,2}S_{3,4}S_{2,4}S_{1,4})+H(S_{1,4}S_{2,4}S_{3,4}S_{2,3}S_{3,2})\nonumber\\
\stackrel{(\ref{eqn:regeneratingencoding})}{=}&H(S_{3,1}S_{2,1}W_4)+H(S_{1,4}S_{2,4}S_{3,2}S_{3,4}W_3)+5B\nonumber\\
&-2H(S_{3,2}S_{3,4}S_{2,4}S_{1,4})+H(S_{1,4}S_{2,4}S_{3,4}S_{2,3}S_{3,2})\nonumber\\
=&H(S_{3,1}S_{2,1}W_4)+H(W_3|S_{1,4}S_{2,4}S_{3,2}S_{3,4})+5B\nonumber\\
&+H(S_{2,3}|S_{1,4}S_{2,4}S_{3,2}S_{3,4})\nonumber\\
\geq& H(S_{3,1}S_{2,1}W_4)+5B+H(W_3S_{2,3}|S_{1,4}S_{2,4}S_{3,4}S_{3,2})\label{eqn:lastbutone}
\end{align}
By writing
\begin{align*}
&H(S_{3,1}S_{2,1}W_4)\stackrel{(s)}{=}H(S_{1,4}S_{3,4}W_2)\stackrel{(\ref{eqn:regeneratingencoding})}{=}H(W_2S_{1,4}S_{3,4}S_{2,3}S_{2,4})\nonumber\\
&=H(W_2|S_{1,4}S_{3,4}S_{2,3}S_{2,4})+H(S_{1,4}S_{3,4}S_{2,3}S_{2,4})
\end{align*}
and
\begin{align*}
&H(W_3S_{2,3}|S_{1,4}S_{2,4}S_{3,4}S_{3,2})\nonumber\\
=&H(W_3S_{2,3}S_{1,4}S_{2,4}S_{3,4}S_{3,2})-H(S_{1,4}S_{2,4}S_{3,4}S_{3,2})\nonumber\\
\stackrel{(s)}{=}&H(W_3S_{2,3}S_{1,4}S_{2,4}S_{3,4}S_{3,2})-H(S_{1,4}S_{2,4}S_{3,4}S_{2,3})\nonumber\\
=&H(W_3S_{3,2}|S_{2,3}S_{1,4}S_{2,4}S_{3,4})
\end{align*}
we can arrive at
\begin{align}
&H(S_{3,1}S_{2,1}W_4)+H(W_3S_{2,3}|S_{1,4}S_{2,4}S_{3,4}S_{3,2})\nonumber\\
\geq& H(W_2W_3S_{3,2}|S_{2,3}S_{3,4}S_{2,4}S_{1,4})+H(S_{1,4}S_{3,4}S_{2,3}S_{2,4})\nonumber\\
=& H(W_2W_3S_{3,2}S_{2,3}S_{3,4}S_{2,4}S_{1,4})
\stackrel{(\ref{eqn:regenerating},\ref{eqn:totalinfo2})}=B. \label{eqn:lastone}
\end{align}
The proof can be completed by combining (\ref{eqn:lastbutone}) and (\ref{eqn:lastone}). \QED
%

\subsection{The Computer-Aided Proof Approach}

It should be clear at this point that the converse proof is rather difficult to find manually, and this difficulty motivated the investigation of the computer-aided proof (CAP) approach. 

Using the linear programming (LP) bound in \cite{Yeung:book} as  an outer bound to the entropy space, one can potentially find an outer bound for the problem. A direct application is however infeasible, because the number of variables in the LP is exponential in the number of random variables, with an even larger number of  constraints. There are at least 16 random variables, resulting in an LP too large for most solvers. To circumvent this difficulty, the symmetry and other techniques are used to reduce the problem size. Moreover, even when certain $(\bar{\alpha},\bar\beta)$ can be shown to be on the outer bound using this approach, it does not lead to an explicit proof as given above. To achieve this, we further extended the LP approach by embedding a secondary linear optimization problem to yield explicitly the converse proof. The details of this approach, unfortunately, can not be included here due to space constraint. 

\section{Conclusion}
\label{sec:conclusion}

A complete characterization is provided for the rate region of the $(4,3,3)$ exact-repair regenerating codes, which shows that the cut-set-based outer bound \cite{Dimakis:10} is in general not (even asymptotically) tight for exact-repair. An explicit binary code construction is provided to show that the given rate region is achievable. It should be noted that the rate region given here remains the same if the codes are required only to have asymptotic zero error probability, instead of zero-error.

As an ongoing work, we are currently investigating the generalization of the results presented here for other $(n,k,d)$ parameters, and have obtained partial results on several more cases, which will be presented in a follow-up work.



\bibliographystyle{IEEEbib}

\end{document}